\documentclass[journal]{IEEEtran}
\usepackage{graphicx}
\usepackage{amsmath}
\usepackage{epsfig}
\usepackage{epstopdf}
\usepackage{bm}
\usepackage{multirow}
\usepackage{array}
\usepackage{booktabs}
\usepackage{array}
\usepackage{tabularx}
 \usepackage{cite}
 \usepackage{balance}

\usepackage{array}
\newcolumntype{L}[1]{>{\raggedright\let\newline\\\arraybackslash\hspace{0pt}}m{#1}}
\newcolumntype{C}[1]{>{\centering\let\newline\\\arraybackslash\hspace{0pt}}m{#1}}
\newcolumntype{R}[1]{>{\raggedleft\let\newline\\\arraybackslash\hspace{0pt}}m{#1}}

\ifCLASSINFOpdf
\else
\fi
\begin{document}
%





\title{Robust Heartbeat Detection from Multimodal Data via CNN-based Generalizable Information Fusion}






%
%
%
\author{B. S. Chandra$^*$, C. S. Sastry and S. Jana%
\thanks{$^*$B. S. Chandra is with the Department of Electrical Engineering, Indian Institute of Technology Hyderabad, India (e-mail: bschandra@iith.ac.in).}
\thanks{C. S. Sastry is with the Department of Mathematics, Indian Institute of Technology Hyderabad, India.}
\thanks{S. Jana is with the Department of Electrical Engineering, Indian Institute of Technology Hyderabad, India.}}

\maketitle


\begin{abstract}

\textit{Objective:} Heartbeat detection remains central to cardiac disease diagnosis and management, and is traditionally performed based on electrocardiogram (ECG). 
To improve robustness and accuracy of detection, especially, in certain critical-care scenarios, the use of additional physiological signals such as arterial blood pressure (BP) has recently been suggested. There, estimation of heartbeat location requires information fusion from multiple signals. However, reported efforts in this direction often obtain multimodal estimates somewhat indirectly, by voting among separately obtained signal-specific intermediate estimates. In contrast, we propose to directly fuse information from multiple signals without requiring intermediate estimates, and thence estimate heartbeat location in a robust manner.
\textit{Method:} We propose as a heartbeat detector, a convolutional neural network (CNN) that learns fused features from multiple physiological signals. This method eliminates the need for hand-picked signal-specific features and ad hoc fusion schemes. Further, being data-driven, the same algorithm learns suitable features from arbitrary set of signals.
\textit{Results:} Using ECG and BP signals of PhysioNet 2014 Challenge database, we obtained a score of 94\%. Further, using two ECG channels of MIT-BIH arrhythmia database, we scored 99.92\%. Both those scores compare favorably with previously reported database-specific results. Also, our detector achieved high accuracy in a variety of clinical conditions. 
\textit{Conclusion:} The proposed CNN-based information fusion (CIF) algorithm is generalizable, robust and efficient in detecting heartbeat location from multiple signals.
\textit{Significance:} In medical signal monitoring systems, our technique would accurately estimate heartbeat locations even when only a subset of channels are reliable.

\end{abstract}


%
\IEEEpeerreviewmaketitle

\section{Introduction}

\begin{figure}[t]
\centering
\begin{tabular}{cc}
\hspace{-3.5mm}
\includegraphics[width=4.5 cm]{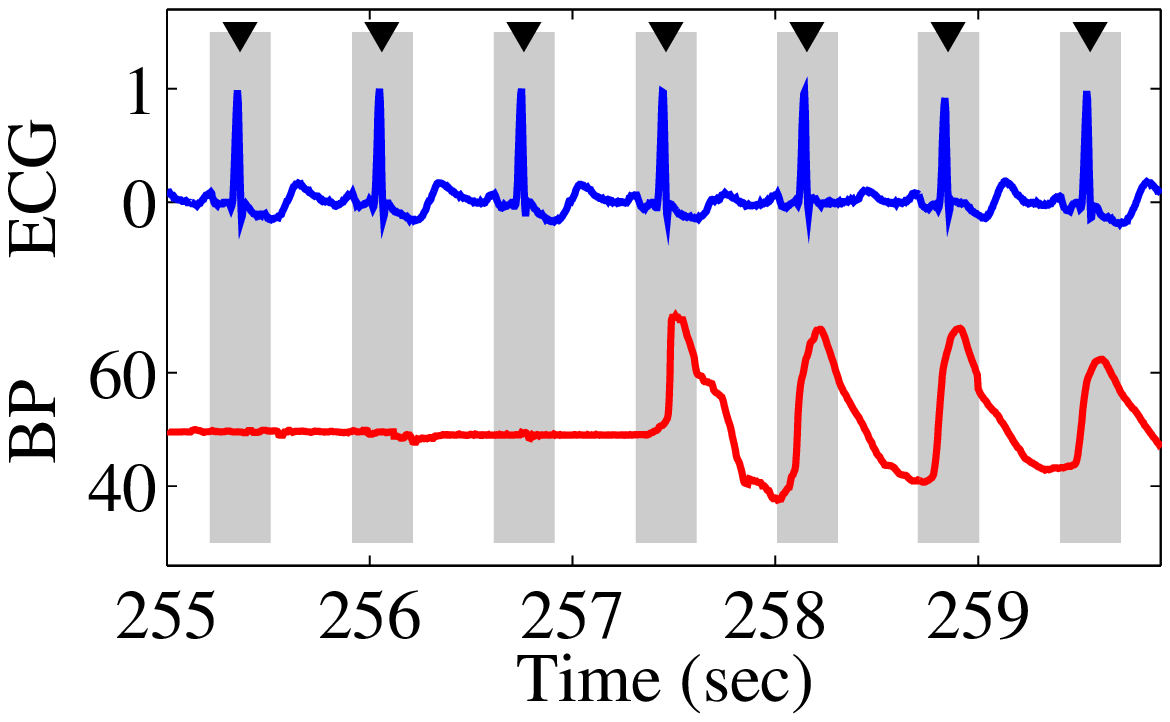} &
\hspace{-5.5mm}
\includegraphics[width=4.5 cm]{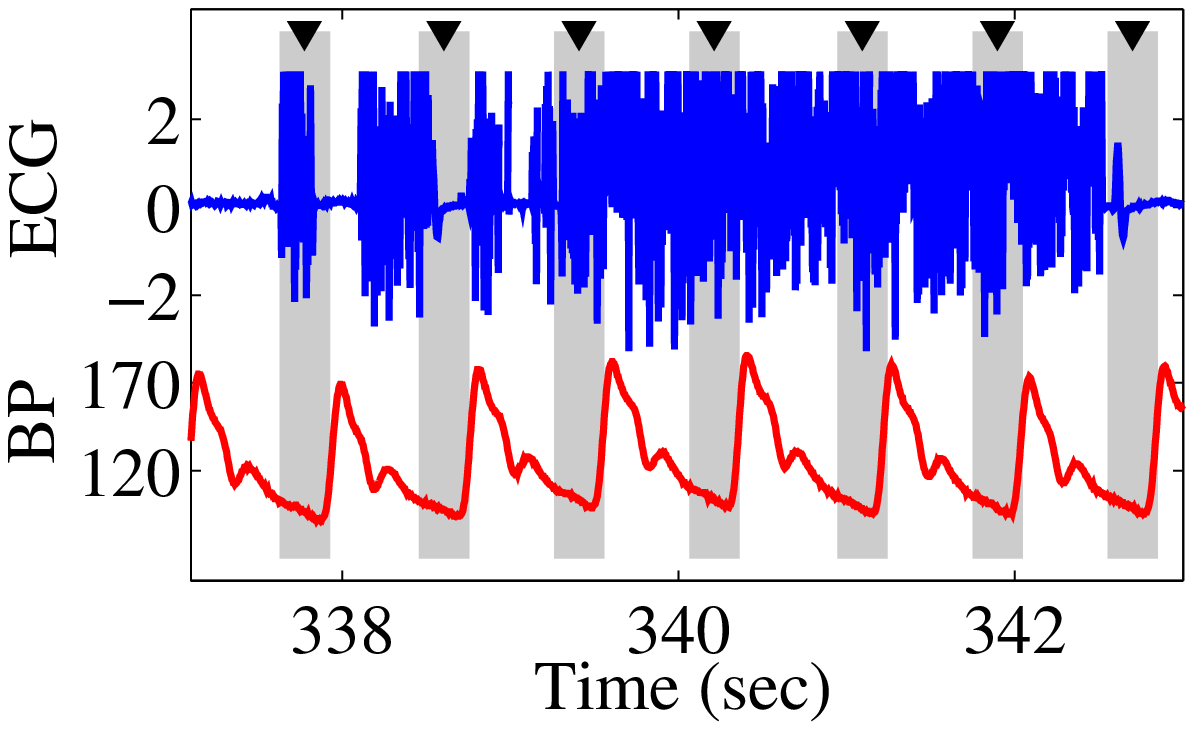} \\ 
\end{tabular}
(a)
\includegraphics[width = 9cm]{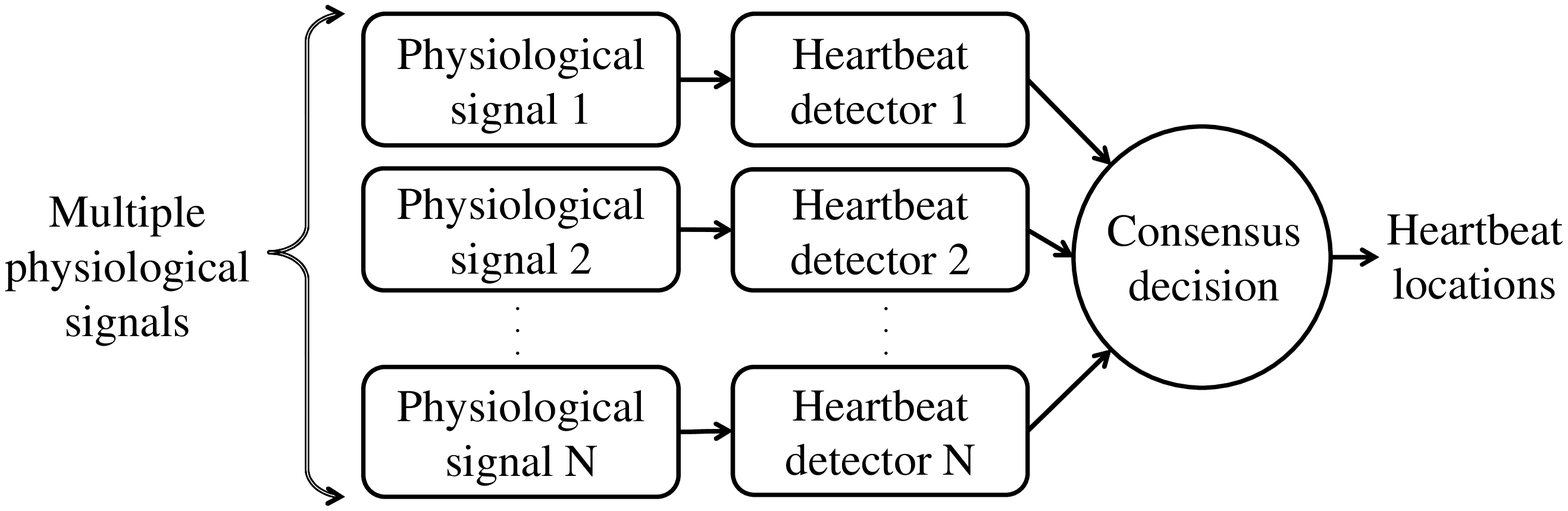} \\
(b) \\
 \quad \\
\includegraphics[width = 9cm]{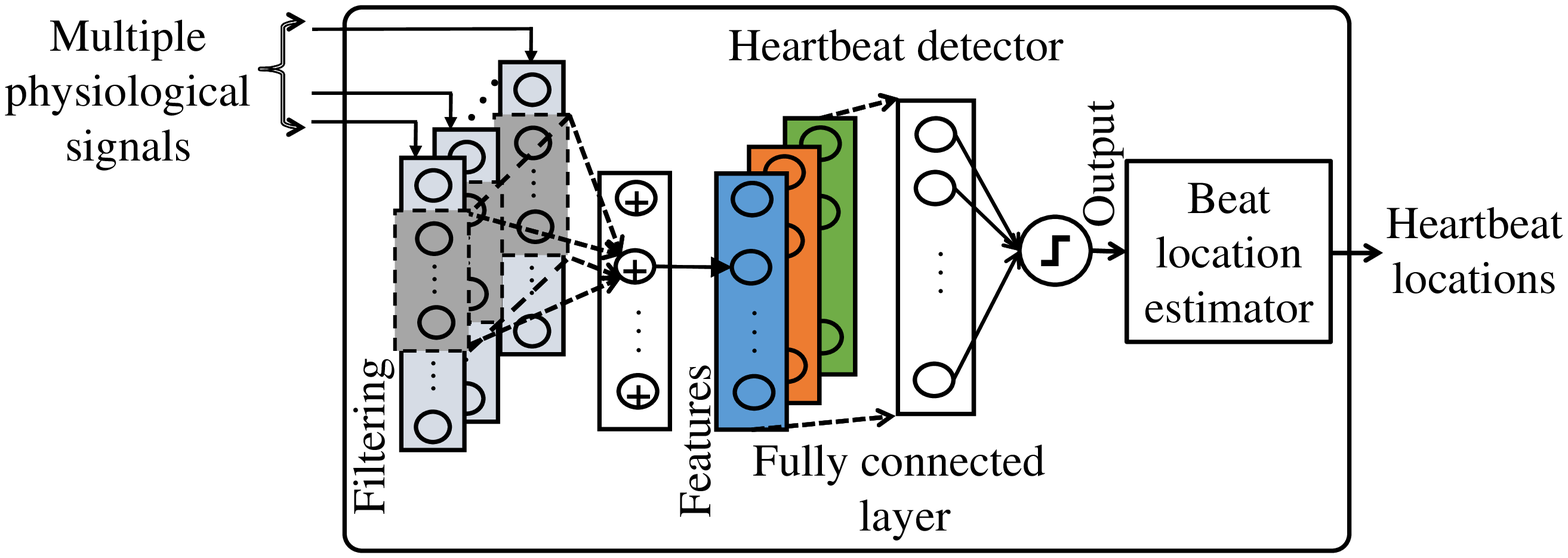} \\
(c) \\

\caption{(a) ECG and BP signals from PhysioNet 2014 Challenge database with annotations at the top \cite{physionet}: left-- segment from record 1071 with clean ECG and missing BP signals, right-- segment from record 2800 with corrupted ECG and clean BP signals; (b) Traditional multimodal heartbeat detector \cite{silva2015robust, ding2016robust, RANKAWAT2017201}; (C) Proposed multimodal detector: CNN-based information fusion (CIF).}
\label{ECG_BP}
\end{figure}

\IEEEPARstart{C}{ardiovascular} diseases (CVDs) are a leading cause of death worldwide \cite{who}, and their management has become a global imperative. In certain scenarios, including intensive care unit (ICU) monitoring, cardiac conditions are continuously assessed for possible worsening. There, automated anomaly detection and alert generation are anticipated to improve timely intervention. In this direction, the first step often involves accurate heartbeat detection based on related physiological signals such as electrocardiogram (ECG) and arterial blood pressure (BP) signals (Figure \ref{ECG_BP}a). Detection based on multiple signals is generally more accurate than that based on individual signals \cite{li2007robust}, because  quality of one signal could be poor,
when other signals maintain high quality. However, a typical multisignal detector \cite{li2007robust, silva2015robust, ding2016robust, RANKAWAT2017201}, simply combining multiple single-signal detections, does not systematically exploit inter-signal correlation. To fill this gap, we propose as a heartbeat detector a convolutional neural network (CNN) that makes a final decision by directly fusing information from multiple signals, without needing intermediate detections.

%




Automated estimation of heartbeat location from ECG signals has received considerable attention. Various techniques including signal differencing, filterbanks, wavelet transform and Hilbert transform followed by threshold comparison have been reported \cite{hamilton1986quantitative, kohler2002principles, christov2004real,  zhang2009qrs,ghaffari2009robust, ghaffari2010segmentation, manikandan2012novel}. To fulfill real-time requirements, adaptive thresholding has been employed \cite{arzeno2008analysis, gutierrez2015novel, ding2014efficient}. Not surprisingly, detection performance is generally improved when multichannel ECG data are used instead of single channel \cite{afonso1999ecg, christov2004real, pangerc2015robust}. Simultaneously, attempts have also been made to estimate heartbeat locations from BP signal \cite{zong2003open}. In general, the aforementioned techniques exploit temporal dependency in each signal. Recently, detection algorithms combining information from ECG and BP signals have been reported \cite{li2007robust, silva2015robust}. Multimodal detectors generally function as shown in Figure \ref{ECG_BP}b \cite{silva2015robust, ding2016robust, RANKAWAT2017201}. An intermediate estimate of heartbeat location is obtained from each signal using hand-picked signal-specific features. Based on such intermediate estimates, a consensus decision is heuristically made, albeit without directly considering inter-signal dependency. An advanced method following this paradigm considers upto three channels each of ECG and blood pressure signals \cite{pangerc2015robust}.
However, exploitation of inter-signal dependency assumes significance because various clinical conditions (e.g., arrhythmias and morphological anomalies) and non-clinical conditions (e.g., noise and pacemakers) affect different signal modes differently. 
Noting this, a recent work employs a hidden semi-Markov model (HSMM) to statistically relate ECG and BP features \cite{pimentel2015heart}. While this method desirably accounts for temporal dependency, HSMM could be restrictive as a model, and the said features are still hand picked for specific signals and not generalizable. In contrast, we propose as a heartbeat detector a CNN, shown in Figure 1c, which learns suitable filters to extract features, temporally fusing information from multiple signals. Finally, a fully connected network maps these features to possible heartbeat locations. As the filter coefficients and the network weights are learned from training data, the proposed CNN-based information fusion (CIF) algorithm generalizes to arbitrary set of signals. 


Generally, a detector performance is evaluated on a public database, which is divided into two sets, used for training and testing. Notionally, the algorithm should be tuned based on the training set, and the performance be recorded on the test set. However, this framework would mimic the real life, only when the detector operates on unforeseen but representative data. 
In this vein, the ``PhysioNet 2014 Challenge" (hereafter ``Challenge") made an extensive training dataset of multiple signals (including ECG and BP) publicly available, while keeping the representative test set hidden. The Challenge held a time-bound competition among detectors, and identified the leaders reporting scores up to 93.64\% \cite{silva2015robust, pangerc2015robust}. Fortunately, the portal continues to score submitted algorithms \cite{ding2016robust, RANKAWAT2017201}, 
and the proposed CIF algorithm achieved a score of 94\%, which compares favorably with hitherto reported scores, indicating its practical significance.
Further, as alluded earlier, our CIF algorithm generalizes to an arbitrary set of physiological signals. To illustrate this, we moved to two channels of ECG. However, in absence of a scoring service that uses undisclosed test data, we turned to MIT-BIH arrhythmia database, based on which performance of a number of existing algorithms has already been reported \cite{hamilton1986quantitative, kohler2002principles, afonso1999ecg, christov2004real, arzeno2008analysis, zhang2009qrs, ghaffari2009robust, manikandan2012novel, gutierrez2015novel, ding2014efficient, pangerc2015robust, ghaffari2010segmentation}. As the test data are not hidden, a performance index here may not directly represent the anticipated real-life accuracy, although performance comparison can still be illuminating. Adopting the Challenge scoring rule, and respectively using only one ECG channel and both the channels, our CIF algorithm achieved scores of 99.89\% and 99.92\%, which ranked among the highest reported so far. Here, we hasten to add that multi-database evaluation has been attempted before. Specifically, a previously mentioned algorithm operating on upto three channels each of ECG and blood pressure signals has been evaluated on five databases, including the Challenge and MIT/BIH databases \cite{pangerc2015robust}. However, as that algorithm depends on features specific to ECG and blood pressure signals, its applicability is accordingly restricted to records consisting of the associated signals only. In contrast, the proposed algorithm generalizes to arbitrary set of signals and records.

In summary, our CIF algorithm for heartbeat detection
\begin{enumerate}
\item fuses information from multiple physiological signals;
\item is data-driven and generalizable; 
\item is efficient and robust for different sets of signals;
\item performs favorably versus signal-specific methods. 
\end{enumerate}
The rest of the paper is organized as follows. In Section \ref{material}, we describe the databases used and the proposed heartbeat detector. Experimental results are provided in Section \ref{res}. Finally, Section \ref{conc} concludes the paper.

\section{Materials and Methods}
\label{material}

We now formalize the problem of heartbeat detection from multiple signals.
%
%
A number $K$ ($\ge 2$) of physiological signals $x^{(1)}_{t_1:t_2}, x^{(2)}_{t_1:t_2}, \ldots, x^{(K)}_{t_1:t_2}$ from a subject are simultaneously recorded from time $t_1$ to time $t_2$. Our task is to estimate heartbeat locations $\tau_1, \tau_2, \ldots, \tau_m$ ($m$ denoting the number of heartbeats) from those signals. Next we describe the databases, outline our CIF heartbeat detector, and present the performance evaluation strategy.

\subsection{Databases}

We make use of PhysioNet 2014 challenge and MIT-BIH arrhythmia databases for performance evaluation \cite{silva2015robust, physionet}. In each database, a patient record is annotated with consensus heartbeat locations marked by two or more cardiologists. In particular, location of the R-peak of an ECG is taken as the heartbeat location. Specific details of the databases are described below.



\subsubsection{Challenge database} The Challenge aims at benchmarking heartbeat detectors operating on multimodal data including ECG and BP \cite{silva2015robust}. The publicly available training set consists of 200 patient records, while the hidden test set has 210 records. Signals from those records have varying duration and sampling frequency varying between 250 and 360 Hz. Also available is a public portal that evaluates and benchmarks the performance of submitted algorithms. Specifically, a score is assigned based on the classification accuracy on 
the entire database consisting of 410 records including training and hidden test data.

\subsubsection{MIT-BIH arrhythmia database} \label{mitbih} 
The MIT-BIH arrhythmia database, often used for evaluating heartbeat detection  \cite{hamilton1986quantitative, kohler2002principles, afonso1999ecg, christov2004real, arzeno2008analysis, zhang2009qrs, ghaffari2009robust, manikandan2012novel, gutierrez2015novel, ding2014efficient, ghaffari2010segmentation, pangerc2015robust} and related algorithms \cite{kiranyaz2016real}, consists of 48 public records, each comprising half-hour two-channel ECG signals sampled at 360Hz. Of those, 25 records, numbered 200-234, contain less common but clinically significant cardiac abnormalities, which we used for training. 
As customary \cite{hamilton1986quantitative, kohler2002principles, afonso1999ecg, christov2004real, arzeno2008analysis, zhang2009qrs, ghaffari2009robust, manikandan2012novel, gutierrez2015novel, ding2014efficient, ghaffari2010segmentation, pangerc2015robust}, the algorithmic performance was evaluated on the entire database consisting of an additional 23 records, numbered 100-124, containing normal beats and a representative set of routine arrhythmias. To maintain consistency, we followed the Challenge rules for final scoring.

\begin{figure}[t]
\centering
\includegraphics[width=8 cm]{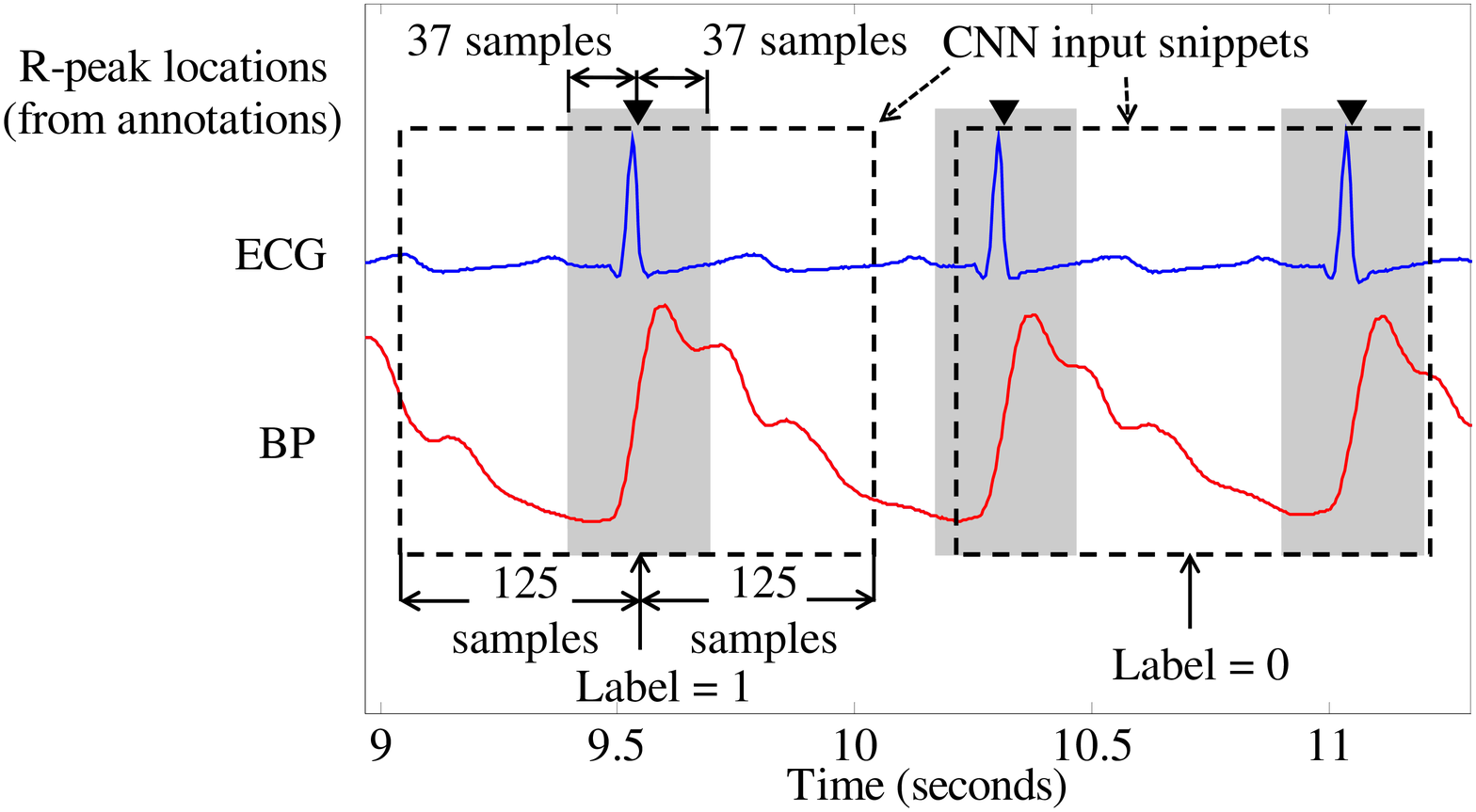}\\
(a) \\
\includegraphics[width=8.5 cm]{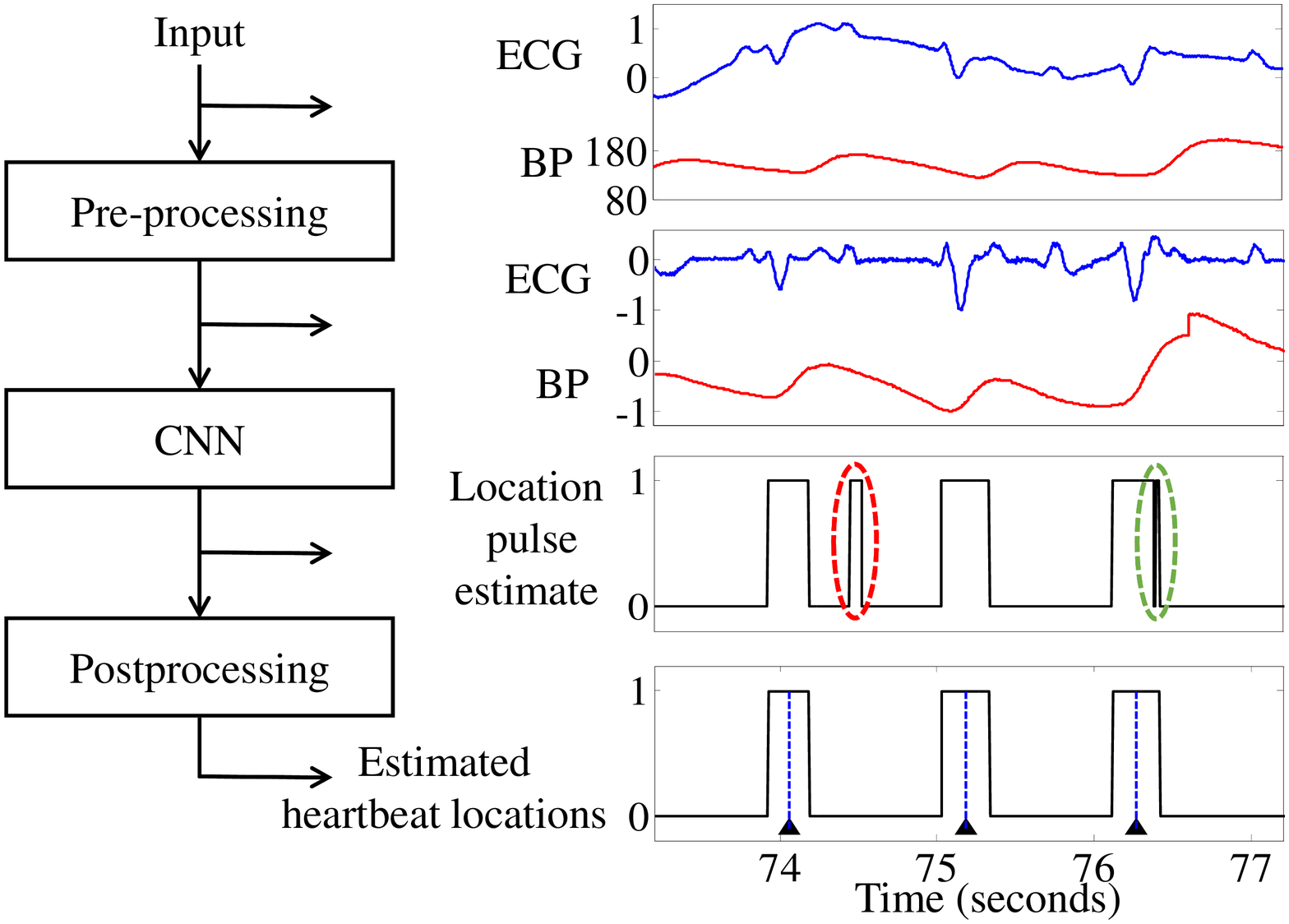} \\
\vspace*{-12pt}
(b)
\caption{(a) CNN input snippets, labeled based on location pulses (shaded gray); (b)
Block diagram of the proposed algorithm with intermediate outputs.}
\label{BD}
\end{figure}




%

\subsection{Detection based on CNN-based Information Fusion (CIF)}

While localizing heartbeats, an estimated location is deemed accurate if it is no further than $150$ms from the corresponding annotated location, in accordance with ANSI/AAMI EC38 and EC57 standards \cite{AAMI}. Thus, a $300$ms location pulse can be formed around the annotated location, and an accurate location estimate falls within such a pulse. As sampling rate varies within as well as across databases, for our analysis, all signals were up or downsampled to a standard rate of $250$Hz. At this rate, the location pulse spanned 75 samples, depicted by gray shade in Figures
\ref{ECG_BP}a, \ref{BD}a, \ref{multimodal}a-d and \ref{MITBIHdata}a-f.

To systematically fuse information from multiple signals, we made use of CNNs. Recall that a CNN adds convolution layers between the input layer and (the hidden as well as) the output layer of a fully-connected neural network \cite{cnn2}. We restricted to one convolution layer consisting of linear filters, whose outputs were passed through a sigmoid nonlinearity onto the fully connected network (see Figure \ref{ECG_BP}c). Assuming $K$ signals, each aforementioned filter consists of $K$ component 1D filters, each with a common length $L$ operating on a specific signal, such that the overall filter output (which is nothing but the desired feature set) is obtained by summing component filter outputs. Formally, assume $p$ filters altogether, and denote by $\{h_l^{(k,j)}\}_{l=0}^{L-1}$ the coefficients of the $k$-th ($1\le k\le K$) 1D channel of the $j$-th ($1\le j\le p$) filter. The corresponding 1D filter output is given by $y_n^{(k,j)}=\sum_{l=0}^{L-1} x_{n-l}^{(k)}h_l^{(k,j)}$. Hence, the overall output of the $j$-th filter is given by $y_n^{(j)}= \sum_{k=1}^{K} y_n^{(k,j)}$. The feature set consists of $y_n^{(j)}$, where $j=1,2,...,p$ and $n$ varies as follows. At a time, we consider a window of length $M$ of signal vector $x_{t:t+M-1}^{(k)}$ for the $k$-th channel with some start time $t$. Further, we avoid zero-padding, so that the $j$-th filter output $y_n^{(j)}$ is obtained only for $n=t+L-1,...,t+M-1$ amounting to $M-L+1$ samples. Thus the length of the feature vector produced by all $p$ filters equals $p(M-L+1)$. Interestingly absence of zero-padding allows a short filter to produce a larger number of outputs than a long filter. The aforementioned filter coefficients as well as network weights were optimized using multimodal training data such that the optimized values essentially capture inter-signal as well as temporal dependence. The said optimization was performed using back propagation algorithm under the cross entropy cost function \cite{cnn2}.

Our CNN input comprises of snippets of $M=251$ samples (duration $1.004$s) each of constituent signals. Each snippet is assigned a label as shown in Figure \ref{BD}a. Specifically, if the middle sample of any snippet falls within any location pulse, the said snippet is labeled 1. It is labeled 0, otherwise. For training, we considered successive snippets with an overlap of $150$ samples (duration of about $0.6$s). As expected, we obtained more training snippets with label 0 than those with label 1 in the process. Subsequently, we maintained class balance by removing randomly chosen snippets with label 0, as required. Alongside original snippets, we also populate the training dataset with modified snippets with only one component signal intact and the rest zeroed so as to enable heartbeat detection even when only one signal is available. The overall training set consists of 551154 snippets for the Challenge database, and 451050 snippets for the MIT-BIH arrhythmia database.



\begin{table*}[ht!]
\centering
\caption{Performance of various CNN architectures (boldface indicates optimized values)}
\label{tab:arch}

\begin{tabular}{|l|cccc|cccc|cccc|>{\centering\arraybackslash}p{0.8cm}>{\centering\arraybackslash}p{0.8cm}>{\centering\arraybackslash}p{0.8cm}|}
\hline
                                 & \multicolumn{12}{c|}{No. of convolution layers = 1, No. of hidden layers = {\bf 0}}  
                                 
                                 & \multicolumn{3}{c|}
                                 {\begin{tabular}[c]{@{}c@{}}
                                No. of convolution layer = 1 \\
                                No. of filters = 2 \\
                                Filter length ($L$) = 20 \end{tabular}} \\  \cline{2-13}
                        Architecture          
                        & \multicolumn{4}{c|}{No. of filters = 1}                                                             & \multicolumn{4}{c|}{No. of filters = {\bf 2}}                                                             & \multicolumn{4}{c|}{No. of filters = 3}                                                             &  \multicolumn{3}{c|}{No. of hidden layers = 1}                                         \\ \cline{2-16} 
                                 & \multicolumn{4}{c|}{Filter length ($L$)}                                                                  & \multicolumn{4}{c|}{Filter length ($L$)}                                                                  & \multicolumn{4}{c|}{Filter length ($L$)}                                                                  &  \multicolumn{3}{c|}{No. of hidden nodes}                                                             \\ \cline{2-16} 
                                 & \multicolumn{1}{c|}{1} & \multicolumn{1}{c|}{20} & \multicolumn{1}{c|}{150} & \multicolumn{1}{c|}{250} & 
                                 \multicolumn{1}{c|}{1} & 
                                 \multicolumn{1}{c|}{{\bf 20}} & \multicolumn{1}{c|}{150} & \multicolumn{1}{c|}{250} & 
                                 \multicolumn{1}{c|}{1} & 
                                 \multicolumn{1}{c|}{20} & \multicolumn{1}{c|}{150} & \multicolumn{1}{c|}{250} & \multicolumn{1}{>{\centering\arraybackslash}p{0.8cm}|}{100} & \multicolumn{1}{>{\centering\arraybackslash}p{0.8cm}|}{200} &                                  \multicolumn{1}{>{\centering\arraybackslash}p{0.8cm}|}{500}         \\ \hline
$Se_{avg}$ (\%)    & 86.2                   & 96.6                    & 95.7                     & 86.3                     & 90.8                   & {96.9}                    & 95.9                     & 88.2                     & 92.5                   & 95.9                    & 95.8                     & 88.1                     & 92.7                              & 93.2                              & 92.1                             \\ 

$PPV_{avg}$ (\%)   & 92.0                   & {97.7}                    & 97.5                     & 91.0                     & 93.7                   & 97.5                    & 97.2                     & 92.2                     & 93.7                   & 97.5                    & 97.2                     & 92.2                     & 95.5                              & 95.6                              & 95.1                             \\ 

$Se_{gross}$ (\%)  & 85.6                   & 96.3                    & 95.6                     & 86.1                     & 90.5                   & {96.7}                    & 95.3                     & 88.0                     & 91.9                   & 95.3                    & 95.3                     & 87.6                     & 92.0                              & 92.6                              & 91.1                             \\ 

$PPV_{gross}$ (\%) & 93.3                   &  {97.8}                    & 97.8                      & 93.6                     & 95.0                   & 97.7                    & 97.4                     & 94.4                     & 96.2                   & 98.1                    & 97.4                     & 93.2                     & 96.2                              & 96.3                              & 95.8                             \\ 

Score (\%)      & 89.3                   & 97.1                    & 96.7                     & 89.2                     & 92.5                   & {\bf 97.2}                    & 96.4                     & 90.7                     & 93.6                   & 96.7                    & 96.4                     & 90.3                     & 94.1                              & 94.4                              & 93.5                             \\ 


\multicolumn{1}{|l|}{Exec. time (ms/beat)}       & 3.1                  & 3.0                   & 3.1               & 3.2                & 3.6                  & 3.6                   & 3.6                  & 3.6                   & 3.8                  & 3.7                   & 3.8                    & 3.9                    & 51.5                           & 51.5                           & 52.4                           \\  
\hline

\end{tabular}
\end{table*}

\begin{figure*}[ht]
\centering
\begin{tabular}{ccccc}
\hspace{-0.5cm} \includegraphics[width=55mm]{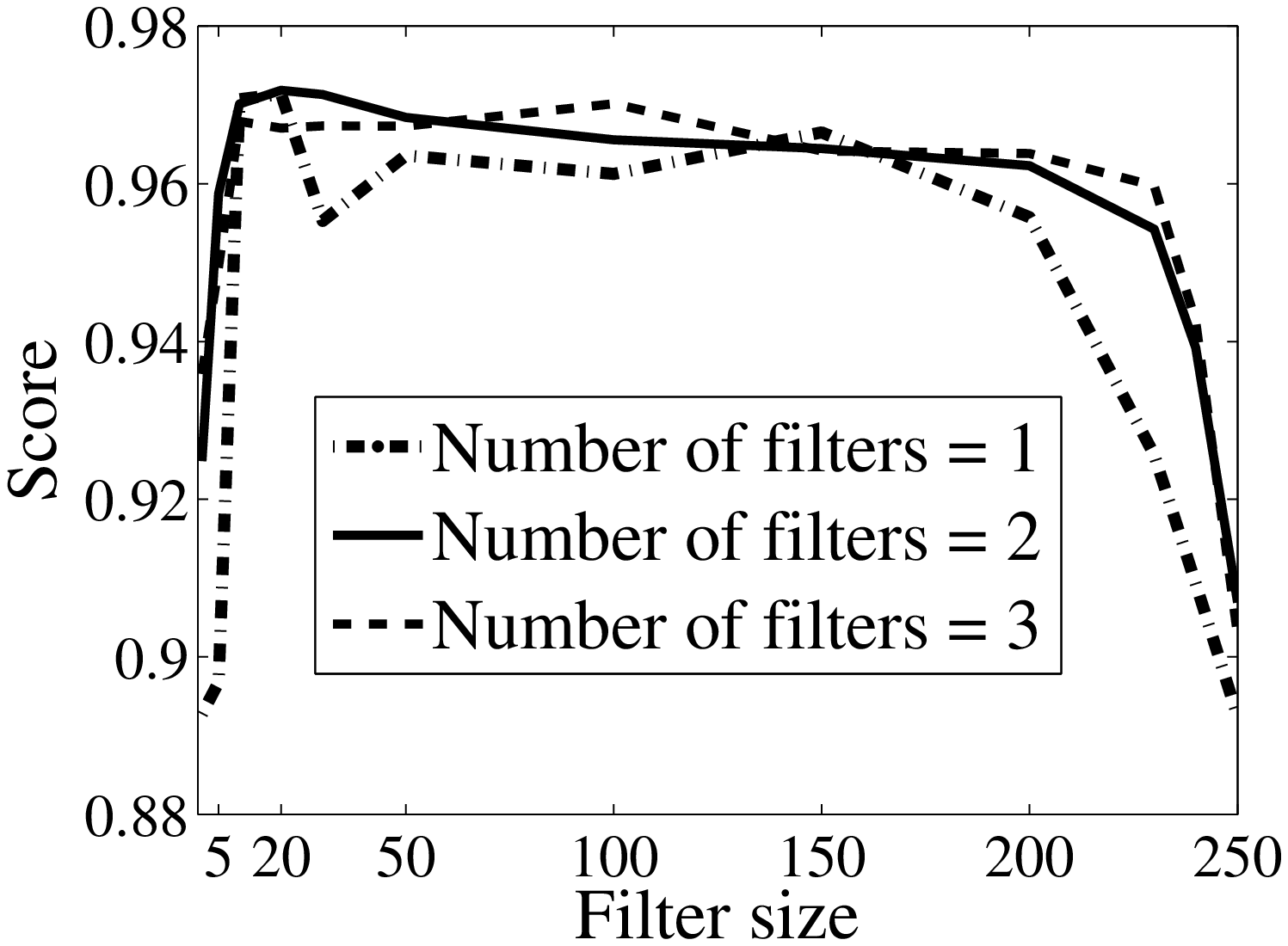} & \hspace{-0.25cm} \includegraphics[width=30mm]{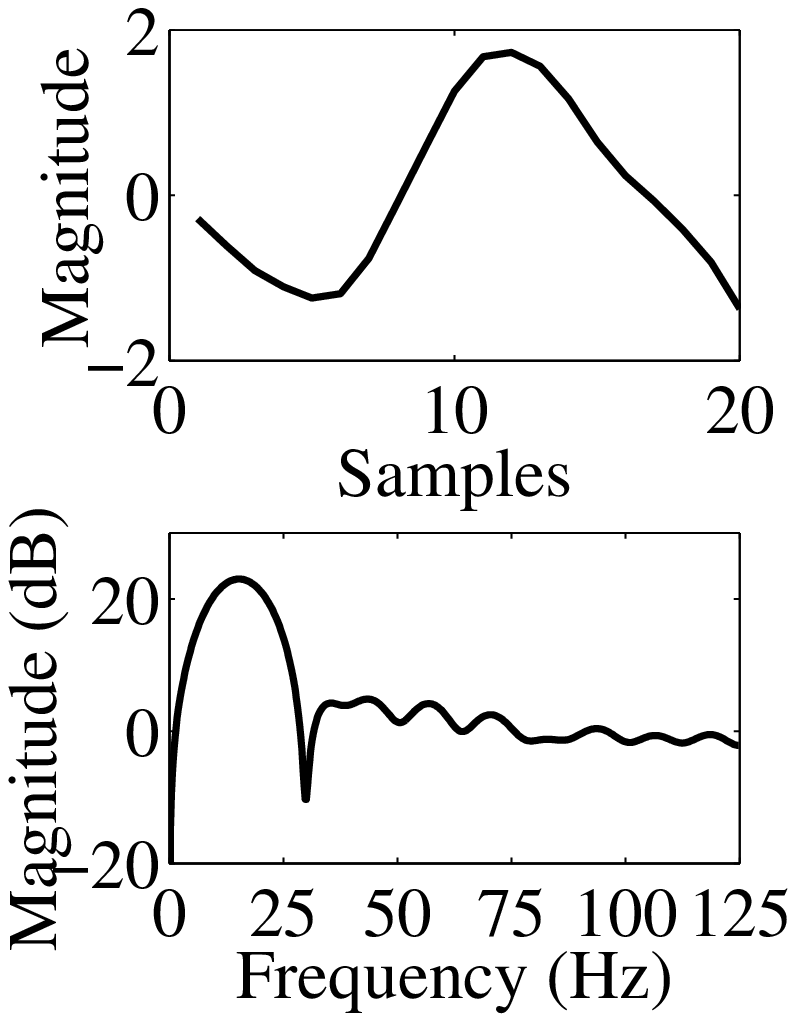}& \hspace{-0.5cm} \includegraphics[width=30mm]{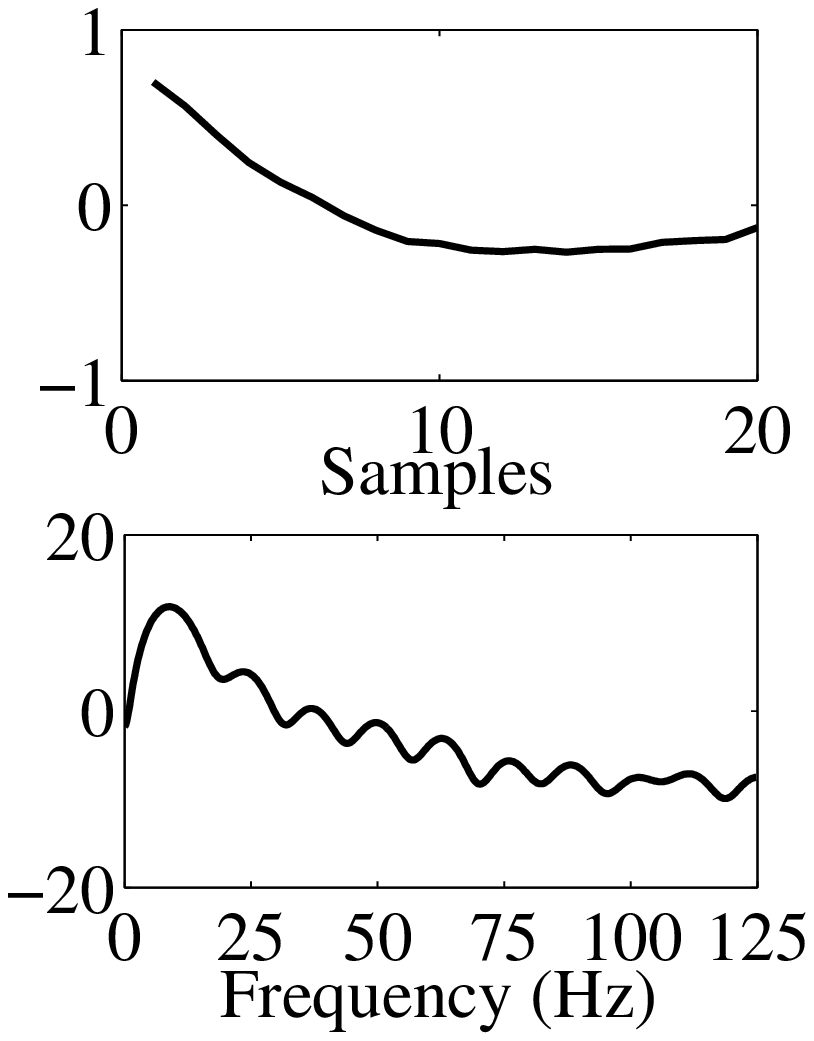}& \hspace{-0.5cm} \includegraphics[width=30mm]{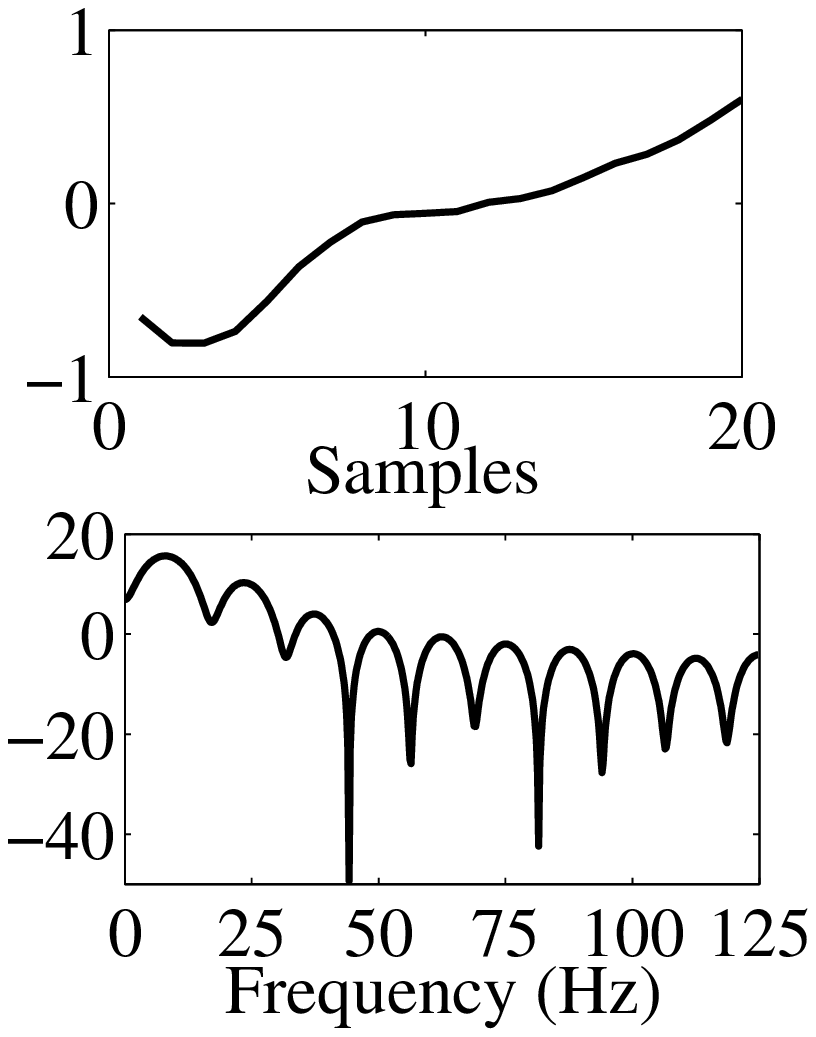}& \hspace{-0.5cm} 
\includegraphics[width=30mm]{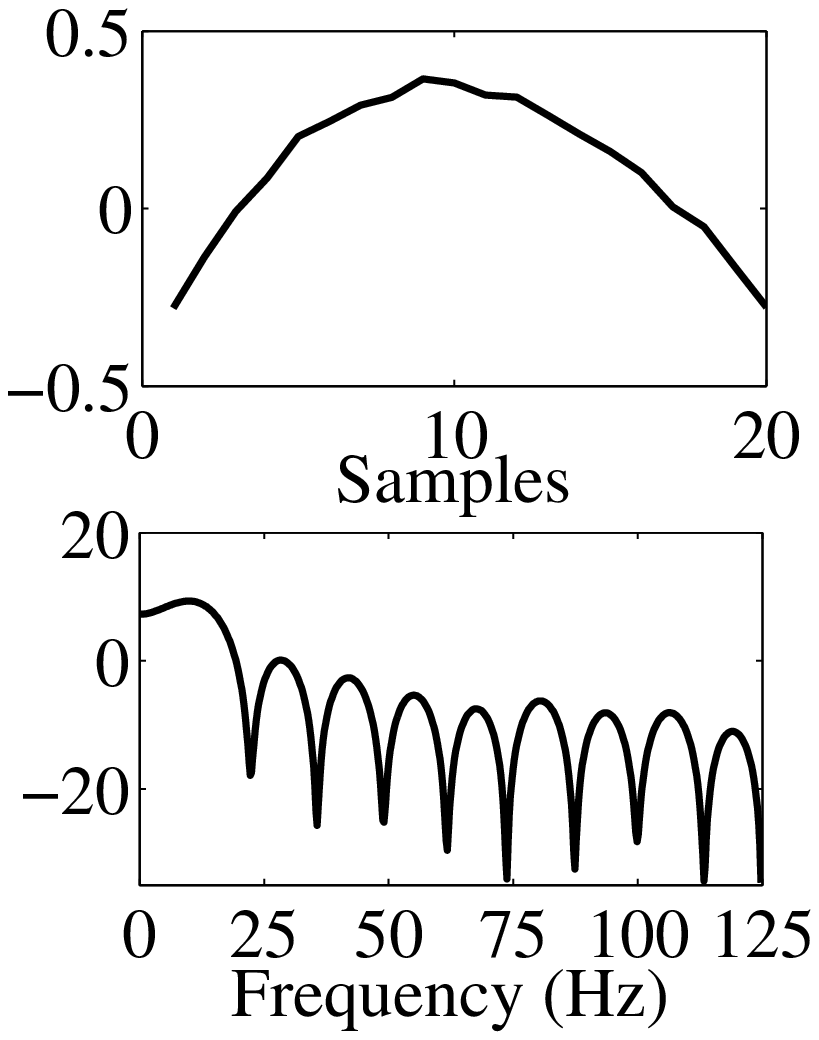} \\
(a) & (b) & (c) & (d) & (e)\\ 
\end{tabular}•
\caption{(a) Performance variation with filter size and number of filters (peak score with two filters of length 20); Learned filters (top) and corresponding frequency response (bottom) for (b) ECG channel in filter-1, (c) BP channel in filter-1, (d) ECG channel in filter-2, (e) BP channel in filter-2.}
\label{filters}
\end{figure*}

To use CIF efficiently, we preprocessed the data as follows. Baseline wander, common in ECG signals, was removed by passing the signal through a cascade of median filters of length 50 and 150 (respective duration $200$ms and $600$ms) \cite{de2004automatic}. Next, range of each signal was normalized to [-1,1]. Specifically, each non-overlapping window of 500 samples (duration $2$s) was scaled by the maximum magnitude. Finally, time synchronization was ensured. For example, in the Challenge database, we compensated for the lag between the ECG and BP signals, which others had found to be around 200ms \cite{johnson2015multimodal}.
We optimized CNN architecture and weights at the training and validation phase. To this end, we used $k$(integer)-fold randomized cross validation on the training set \cite{arlot2010survey}. 

Test snippets were generated from a test record by moving a window of length 251 one sample at a time. Such snippets are presented sequentially to the trained and optimized CNN, and the output sequence was viewed as location pulse estimate. We propose a two-step postprocessing for such initial estimate to improve accuracy. Firstly, we considered inverted pulses, i.e., intervals with label 0 preceded and succeeded by regions with label 1. Very short inverted pulses lasting less than 3 samples (duration 10ms) arise likely due to noise. Such an inverted pulse can be located within the region circled in green in Figure \ref{BD}b. We propose to remove each such spurious inverted pulse by switching the label from 0 to 1. Secondly, recall that a bona fide location pulse is 75 samples long and separation between two heartbeats is generally greater than 100 samples (corresponding to 150 beats/min). Therefore, an estimated pulse lasting less than 50 samples (duration 200ms) is likely spurious. For an example, refer to the region circled in red in Figure \ref{BD}b. We propose to remove such a pulse by switching the label from 1 to 0. The resulting pulses were taken as the final estimates of location pulses, and the midpoints of such pulse estimates were taken as estimated heartbeat locations.

\subsection{Performance evaluation}
For performance evaluation using either database, we adopted the Challenge evaluation scheme, which is described below. Recall that a location estimate was deemed correct if it fell within $150$ms of the annotated location \cite{AAMI}. For the $i$-th record, we counted the numbers of correctly detected beats (true positives, $TP_i$), missed beats (false negatives, $FN_i$) and spurious detections (false positives, $FP_i$), and hence calculated sensitivity $Se_{i} = TP_i/(TP_i+FN_i)$ and positive predictive value (PPV) $PPV_{i} = TP_i/(TP_i+FP_i)$. Subsequently, recordwise average sensitivity $Se_{avg} = \sum_{i=1}^{N}Se_i/N $ and average PPV $PPV_{avg} = \sum_{i=1}^{N}PPV_i/N $ were computed, where $N$ denotes the total number of records. We also calculated gross sensitivity $Se_{gross} = TP/(TP+FN)$ and gross PPV $PPV_{gross} = TP/(TP+FP)$, where $TP=\sum_{i=1}^N TP_i$, $FN=\sum_{i=1}^N FN_i$ and $FP=\sum_{i=1}^N FP_i$. Finally, the algorithm was assigned an overall performance score of $(Se_{avg}+PPV_{avg}+Se_{gross}+PPV_{gross})/4$.

\section{Experimental Results}
\label{res}

Now we present the experimental results and compare the performance of the proposed CIF algorithm with existing heartbeat detectors. To this end, we made use of the ECG and BP signals of PhysioNet 2014 Challenge database and later two ECG channels of MIT-BIH arrhythmia database.

\begin{table}[t!]
\centering
\caption{Performance comparison on PhysioNet 2014 challenge hidden dataset (highest values are boldfaced)}
\label{challengePerf}
\begin{tabular}{@{}p{3cm}p{0.5cm}p{0.75cm}p{0.75cm}p{0.9cm}c}
\toprule
Algorithm & $Se_{avg}$ & $PPV_{avg}$ & $Se_{gross}$ & $PPV_{gross}$ & Score \\ \toprule

Galeotti {\em et al} 2015 \cite{galeotti2015robust} & 91.08 & 86.96 & 92.74 & 87.37 & 89.53 \\

DeCooman {\em et al} 2015 \cite{de2014heart} &89.59  &89.62  &90.74  &90.15  & 90.02 \\

Antink {\em et al} 2015 \cite{antink2015detection} & 89.40 & 90.70 & 91.02 & 91.87 & 90.70 \\

Johnson {\em et al} 2015 \cite{johnson2015multimodal} & 92.61 & 89.03 & 95.07 & 89.30 & 91.50 \\

Pangerc {\em et al} 2015 \cite{pangerc2015robust} & {\bf 93.86} & 91.57 &  {\bf 95.65} & 93.48 & 93.64 \\

Ding {\em et al} 2016 \cite{ding2016robust}  & 89.36 & 87.15 &  91.06 & 87.09 & 88.66 \\

Rankawat {\em et al} 2016 \cite{RANKAWAT2017201}  & 91.60 & 88.85 &  92.74 & 90.39 & 90.89 \\

{\bf Proposed CIF algorithm 
} & 92.85 & {\bf 94.29} & 93.41 & {\bf 95.47} & {\bf 94.00} \\

CIF algorithm (only ECG) & 85.72	 & 89.70  & 87.07	& 91.20 & 88.42 \\

CIF algorithm (only BP) & 	67.00	& 88.66 &  70.11 & 94.26 & 80.00
\\ \hline
\end{tabular}
\end{table}

\begin{figure*}[t!]
\centering
\includegraphics[width=17.5cm]{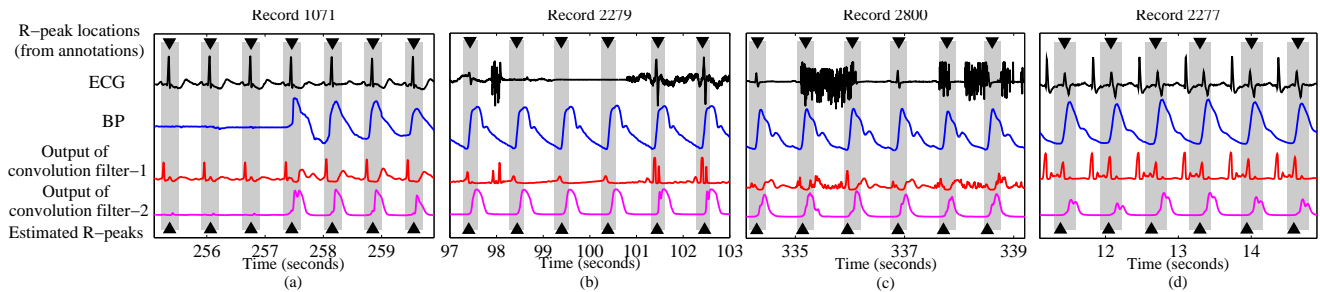} 
\caption{R-peak detection with (a) missing data in BP channel; (b) missing data in ECG channel; (c) noisy ECG channel; (d) paced beats.}
\label{multimodal}
\end{figure*}

\begin{table*}[t]
\centering
\caption{Performance comparison with other methods on MIT-BIH arrhythmia dataset (Superior value in boldface)}
\label{MITBIHcomp}
\begin{tabular}{@{}l@{}cccccccccc@{}}
\toprule
Algorithm      & No. of channels   & Total beats & TP     & FP        & FN       & $Se_{gross}$     & $PPV_{gross}$     & $Se_{avg}$   & $PPV_{avg}$     &  Score \\ \midrule
Hamilton {\it et al.} (1986)  \cite{hamilton1986quantitative}     &  1   & 109267      &   108927     & 248       & 340      & 99.69  & 99.77  & 99.70 & 99.76   &  99.73\\
Afonso {\it et al.} (1999) \cite{afonso1999ecg}    & 2   & 91283        &     90909   & 406       & 374      & 99.59  & 99.56  &   99.61    &   99.53  & 99.57\\

Christov: Algorithm-1 (2004) \cite{christov2004real}      & 2 & 110050      & 109756 & 215       & 294      & 99.73  &  99.80 &   99.75    &  99.80 	&      99.77 \\
Christov: Algorithm-2 (2004) \cite{christov2004real}       & 2  & 110050      & 109810 & 239       & 240      & 99.78  &  99.78 &  99.80      & 99.78	&     99.79\\

Arzeno {\it et al.} (2008)  \cite{arzeno2008analysis} & 1 & 109453      & 109099 & 405  & 354 &   99.68     &      99.63  &   --    &     --    &	-- \\
Zhang {\it et al.} (2009)       \cite{zhang2009qrs}  	&  1 & 109510      &   109297     & 204       & 213      & 99.80  & 99.81   &    99.80   & 99.81  &	 99.81\\

Ghaffari {\it et al.} (2009)       \cite{ghaffari2009robust}  	& 1 & 109428      &   109327     & 129       & 101      & 99.91  & 99.88   &    --   &  --  &	 --\\

Ghaffari {\it et al.} (2010) \cite{ghaffari2010segmentation}  	& 1 & 109428      &   109367     & 89       & {\bf 61}      & {\bf 99.94}  & 99.91   &    --   &  --  &	 --\\

Manikandan {\it et al.} (2012)  \cite{manikandan2012novel} 	& 1 &     109496        &   109417     & 140       & 79       & 99.93  & 99.87  & 99.93  & 99.86 &  99.90\\

Ding {\it et al.} (2014) \cite{ding2014efficient}         & 1   & 109494      & 109360 &  134       & 73     & 99.93    & 99.88  & 	99.93   &  99.88      &   99.91 \\

Guti{\'e}rrez-Rivas {\it et al.} (2015) \cite{gutierrez2015novel}         & 1   & 109949      & 109447 & 289       & 502      & 99.54 & 99.74 &  --     & 	--  &   --\\

Pangerc {\em et al} (2015) (channel-1) \cite{pangerc2015robust}   & 1   & 109494      & 109369 & 125       & 125      & 99.89 & 99.89 &  --     & 	--  &   --\\

Pangerc {\em et al} (2015) (both channels) \cite{pangerc2015robust}   & 2   & 109494      & 109380 & 92       & 114      & 99.90 & 99.92 &  --     & 	--  &   --\\

Elgendi {\it et al.} (2017) \cite{elgendi2017efficient}         & 1   &  109985 & 109775 & 82  & 247 & 99.78 & 99.93 &  99.78     & 	99.92  &   99.85 \\

CIF algorithm (channel-1)       & 1     & 109494      & 109322 & {\bf 64}       & 172      & 99.84  & {\bf 99.94}   &    99.84   &   {\bf 99.95}  &    99.89   \\

{\bf Proposed CIF algorithm}       & 2    & 109494      & 109422 & 103       & 72      & 99.93  & 99.91   &   {\bf 99.94}  &  99.91 &    {\bf 99.92}   \\ \bottomrule
\end{tabular}
\end{table*}

\begin{table}[t]
\centering
\caption{Recordwise performance on MIT-BIH arrhythmia database}
\label{FullPerf}
\begin{tabular}{@{}rp{0.6cm}r@{}c@{}lp{0.4cm}p{0.4cm}r@{}c@{}lp{0.4cm}p{0.4cm}r@{}c@{}l@{}} \toprule
      &        & \multicolumn{5}{c}{Using only channel-1} & \multicolumn{5}{c}{Using both channels} & \\ \cmidrule(l){3-7} \cmidrule(l){8-12} 
Rec   & Peaks  & FP &$|$& FN    & Se     & PPV    & FP & $|$ & FN   & Se      & PPV  &  FP & $|$ & FN \\ \midrule
100   & 2273   & 0 & $|$ & 1     & 99.96  & 100    & 0 & $|$ & 1    & 99.96   & 100   & $-$ & $|$ & $-$ \\
101   & 1865   & 0 & $|$ & 1     & 99.95  & 100    & 1 & $|$ & 1    & 99.95   & 99.95 & $ \uparrow $ & $|$ & $ -$ \\
102   & 2187   & 0 & $|$ & 0     & 100    & 100    &  1 & $|$ & 1    & 99.95   & 99.95  &  $\uparrow $ & $|$ & $ \uparrow$\\
103   & 2084   & 0 & $|$ & 0     & 100    & 100    &  0 & $|$ & 0    & 100     & 100   & $- $ & $|$ & $ -$ \\
104   & 2229   & 0 & $|$ & 1     & 99.96  & 100    & 0 & $|$ & 0    & 100   & 100   & $- $ & $|$ & $ \downarrow$ \\
105   & 2572   & 16 & $|$ & 22    & 99.14  & 99.38  & 11 & $|$ & 12   & 99.53   & 99.57 & $\downarrow $ & $|$ & $ \downarrow$\\
106   & 2027   & 2 & $|$ & 2     & 99.90  & 99.90  & 2 & $|$ & 2    & 99.90   & 99.90 & $- $ & $|$ & $ -$  \\
107   & 2137   & 1 & $|$ & 1     & 99.95  & 99.95  & 1 & $|$ & 1    & 99.95   & 99.95 & $- $ & $|$ & $ -$  \\
108   & 1763   & 2 & $|$ & 6     & 99.66  & 99.89  & 2 & $|$ & 5    & 99.72   & 99.89 & $- $ & $|$ & $ \downarrow$  \\
109   & 2532   & 0 & $|$ & 0     & 100    & 100    & 1 & $|$ & 0    & 100     & 99.96 & $\uparrow $ & $|$ & $ -$  \\
111   & 2124   & 0 & $|$ & 2     & 99.91  & 100    & 0 & $|$ & 0    & 100     & 100 & $- $ & $|$ & $ \downarrow$    \\
112   & 2539   & 0 & $|$ & 0     & 100    & 100    & 0 & $|$ & 0    & 100     & 100  & $- $ & $|$ & $ -$  \\
113   & 1795   & 0 & $|$ & 1     & 99.94  & 100    & 0 & $|$ & 1    & 99.94   & 100  &  $- $ & $|$ & $-$  \\
114   & 1879   & 2 & $|$ & 0     & 100    & 99.89  & 0 & $|$ & 0    & 100     & 100 &  $\downarrow $ & $|$ & $ -$   \\
115   & 1953   & 0 & $|$ & 0     & 100    & 100    & 6 & $|$ & 1    & 99.95   & 99.69 &  $\uparrow $ & $|$ & $ \uparrow$  \\
116   & 2412   & 0  & $|$ & 20    & 99.17  & 100    & 0 & $|$ & 0    & 100     & 100  &  $- $ & $|$ & $ \downarrow$  \\
117   & 1535   & 0 & $|$ & 0     & 100    & 100    & 0 & $|$ & 0    & 100     & 100  &  $- $ & $|$ & $ -$  \\
118   & 2278   & 1 & $|$ & 0     & 100    & 99.96  & 1 & $|$ & 0    & 100     & 99.96 &  $- $ & $|$ & $ -$ \\
119   & 1987   & 0 & $|$ & 0     & 100    & 100    & 0 & $|$ & 0    & 100     & 100  &  $-$ & $|$ & $ -$  \\
121   & 1863   & 0 & $|$ & 2     & 99.89  & 100    & 2 & $|$ & 0    & 100     & 99.89 &  $\uparrow $ & $|$ & $ \downarrow$ \\
122   & 2476   & 0 & $|$ & 0     & 100    & 100    & 0 & $|$ & 0    & 100     & 100  &  $- $ & $|$ & $ -$  \\
123   & 1518   & 1 & $|$ & 0     & 100    & 99.93  & 1 & $|$ & 0    & 100     & 99.93 &  $- $ & $|$ & $ -$ \\
124   & 1619   & 0 & $|$ & 0     & 100    & 100    & 0 & $|$ & 0    & 100     & 100  &  $- $ & $|$ & $ -$  \\
200   & 2601   & 2 & $|$ & 2     & 99.92  & 99.92  & 2 & $|$ & 2    & 99.92   & 99.92 &  $- $ & $|$ & $ -$  \\
201   & 1963   & 1 & $|$ & 22    & 98.88  & 99.95  & 1 & $|$ & 1    & 99.95   & 99.95 &  $- $ & $|$ & $ \downarrow$  \\
202   & 2136   & 0 & $|$ & 5     & 99.77  & 100    & 1 & $|$ & 0    & 100     & 99.95 &  $\uparrow $ & $|$ & $ \downarrow$ \\
203   & 2958   & 22 & $|$ & 18   & 99.40   & 99.26 & 36 & $|$ & 22    & 99.26  & 98.8 &  $\uparrow $ & $|$ & $ \uparrow$ \\
205   & 2656   & 0 & $|$ & 2     & 99.92  & 100    & 0 & $|$ & 0    & 100     & 100  &  $- $ & $|$ & $ \downarrow$ \\
207   & 1860   & 5 & $|$ & 11    & 99.41  & 99.73  & 11 & $|$ & 6    & 99.68   & 99.41 &  $\uparrow $ & $|$ & $ \downarrow$ \\
208   & 2955   & 1 & $|$ & 12    & 99.59  & 99.97  & 6 & $|$ & 9    & 99.70   & 99.80 &  $\uparrow $ & $|$ & $ \downarrow$ \\
209   & 3005   & 0 & $|$ & 0     & 100    & 100    & 1 & $|$ & 1    & 99.97   & 99.97 &  $\uparrow $ & $|$ & $ \uparrow$ \\
210   & 2650   & 3 & $|$ & 4     & 99.85  & 99.89  & 4 & $|$ & 1    & 99.96   & 99.85 &  $\uparrow $ & $|$ & $ \downarrow$ \\
212   & 2748   & 1 & $|$ & 0     & 100    & 99.96  & 2 & $|$ & 0    & 100     & 99.93  &  $\uparrow $ & $|$ & $ -$ \\
213   & 3251   & 0 & $|$ & 1     & 99.97  & 100    & 0 & $|$ & 1    & 99.97   & 100  &  $- $ & $|$ & $ -$ \\
214   & 2262   & 0 & $|$ & 4     & 99.82  & 100    & 0 & $|$ & 3    & 99.87   & 100  &  $- $ & $|$ & $ \downarrow$  \\
215   & 3363   & 0 & $|$ & 0     & 100    & 100    & 0 & $|$ & 0    & 100     & 100 &  $- $ & $|$ & $ -$  \\
217   & 2208   & 0 & $|$ & 3     & 99.86  & 100    & 0 & $|$ & 0    & 100     & 100  &  $- $ & $|$ & $ \downarrow$  \\
219   & 2154   & 0 & $|$ & 0     & 100    & 100    & 0 & $|$ & 0    & 100     & 100 &  $- $ & $|$ & $ -$   \\
220   & 2048   & 0 & $|$ & 0     & 100    & 100    & 0 & $|$ & 0    & 100     & 100  &  $- $ & $|$ & $ -$  \\
221   & 2427   & 0 & $|$ & 0     & 100    & 100    & 2 & $|$ & 0    & 100     & 99.92 &  $\uparrow $ & $|$ & $ -$ \\
222   & 2483   & 1 & $|$ & 5     & 99.80  & 99.96  & 2 & $|$ & 0    & 100     & 99.92 &  $\uparrow $ & $|$ & $ \downarrow$ \\
223   & 2605   & 0 & $|$ & 0     & 100    & 100    & 1 & $|$ & 0    & 100     & 99.96  &  $\uparrow $ & $|$ & $ -$ \\
228   & 2053   & 3 & $|$ & 18    & 99.12  & 99.85  & 0 & $|$ & 1    & 99.95   & 100  &  $- $ & $|$ & $ \downarrow$ \\
230   & 2256   & 0 & $|$ & 0     & 100    & 100    & 0 & $|$ & 0    & 100     & 100  &  $- $ & $|$ & $ -$  \\
231   & 1571   & 0 & $|$ & 0     & 100    & 100    & 0 & $|$ & 0    & 100     & 100  &  $- $ & $|$ & $ -$  \\
232   & 1780   & 0 & $|$ & 5     & 99.72  & 100    & 5 & $|$ & 0    & 100     & 99.72 &  $\uparrow $ & $|$ & $ \downarrow$ \\
233   & 3079   & 0 & $|$ & 1     & 99.97  & 100    & 0 & $|$ & 0    & 100     & 100  &  $- $ & $|$ & $ \downarrow$  \\

234   & 2753   & 0 & $|$ & 0     & 100    & 100    & 0 & $|$ & 0    & 100     & 100  &  $- $ & $|$ & $ -$  \\
Gross & 109494 & 64 & $|$ & 172   & 99.84  &    99.94    & 103 & $|$ & 72   &   99.93    &   99.91  &  $\uparrow $ & $|$ & $ \downarrow$   \\
Avg   &        &   & &   &      99.84   &   99.95     &      & &      &   99.94      &       99.91 &  &  & \\  \bottomrule
\end{tabular}
\end{table}

\subsection{Results for Challenge Database}
\label{sec:ChRes}

We first optimized the CNN for the Challenge database and evaluated its performance.

\underline{\em CNN Optimization:}
We optimized the CNN architecture as well as weights. For the sake of simplicity, we considered shallow architectures with single convolution and at most one hidden layers. Using two channels, ECG and BP, of the Challenge training set, any CNN architecture with optimized weights was given a 10-fold randomized cross validation score \cite{arlot2010survey}, based on which various configurations were compared. First, considering no hidden layers, the number of filters and the common filter length were optimized. To simplify matters, only one-, two- and three-filter configurations were explored. Also, the filter length was varied from 1 to 250 (see Table \ref{tab:arch} and Figure \ref{filters}a). A short filter scores low because, while outputting many features, it considers only a short temporal span, and does not exploit temporal dependency efficiently. A long filter also scores low because, while observing for a significant duration, has only a few outputs to encode the temporal dependency into. In general, as the filter length increases, scores rise sharply, remain somewhat steady, and then fall sharply for each filter.
Here, the plateau region indicates that a filter length taken from a wide interval performs satisfactorily. The peak scores for one-filter and two-filter cases are both achieved at a filter length of 20 and are nearly equal, while the score for three filters is somewhat less. Of the first two, we picked the latter because the plateau around the peak is relatively flatter, facilitating robust operation. 
Next, using the above two-filter configuration with filter length 20 as reference, a hidden layer was introduced, and the aforementioned score was recorded vis-\`{a}-vis varying number of hidden nodes in anticipation of possible improvement, which did not materialize (Table \ref{tab:arch}). Also, with the introduction of the hidden layer, the execution time per beat, using MATLAB v.2014b run on a desktop computer with an Intel core i7 3.4 GHz 64-bit processor and 16 GB memory, increased by more than an order of magnitude.  So, we persisted with the aforementioned two-filter configuration without hidden layers.

Recall that each filter operates on the two-channel data (ECG and BP) and generates a feature vector that captures both the temporal and inter-signal dependencies. Let us now inspect the optimized filter characteristics in time as well as frequency domain as presented in Figures \ref{filters}b-e. Here the frequency response corresponding to one (ECG or BP) channel of the filter is obtained by zeroing the other (BP or ECG) channel. Observe that the ECG channel of filter-1 learned to pass a 3db frequency band of approximately 9.25--21.25Hz. This passband generally matches the recommended frequency range for detecting QRS complexes \cite{benitez2000new}, and approximates the passband of 10--24Hz of ECG-specific slope- and peak-sensitive filters \cite{pangerc2015robust}. Turning to time domain, the present impulse response bears similarity with Haar-like matched filters and mother wavelets that have been used in representing the shape of QRS complexes \cite{ding2014efficient, ghaffari2010segmentation}. In short, the ECG channel of our learned filter-1 appears to capture much of human intuition. On the other hand, the BP channel of learned filter-1 passes a narrower and lower frequency band of 4.15--14.16Hz with more pronounced sidelobes. Interestingly, this passband does not overlap with the passband of 1.2--3.5Hz of a filter used to detect the steep-slope segment of blood pressure signals \cite{pangerc2015robust}. Thus, while the learned filter-1 appears to corroborate the effectiveness of a slope- and peak-sensitive filter for the ECG channel, it does not for the BP channel. In general, the CNN presents an alternative paradigm, where two filters seem to play complementary roles. Specifically, compared to the ECG channel of filter-1, filter-2 has a passband 3.66--12.38Hz of significantly lower frequencies and more pronounced sidelobes. Turning to the BP channel, filter-2 has lowpass response with a cutoff frequency of 15.88Hz, complementing the bandpass response of filter-1.

\underline{\em Heartbeat detection performance:}
As mentioned earlier, Challenge score reports are generated based on a public training set and a hidden test set, designed to reflect real-life complexities. In Table \ref{challengePerf}, we report sensitivity and PPV values, both gross and average, as well as the overall score obtained by the proposed CIF as well as existing algorithms. Our overall score of 94.00\% is significantly higher than 88.42\% and 80.00\%, respectively obtained using only ECG and only BP, illustrates the benefit of fusion, and compares favorably with hitherto reported scores. 
Importantly, our heartbeat detector was found to be robust to various non-clinical conditions shown in Figures \ref{multimodal}a-c, where one of the BP and the ECG signals is either missing or heavily corrupted. There, filter-1 and filter-2 appeared to prominently encode ECG and BP features, respectively, whose simultaneous use ensured robust detection.
Further, consider paced beats, induced by pace maker implants, which are not genuine heartbeats, but conventional detectors often confuse those with normal beats \cite{johnson2015multimodal}. Desirably, our detector correctly ignored paced beats, as shown in Figure \ref{multimodal}d, as a pace maker leaves no corresponding BP signature.

\subsection{Results for MIT-BIH Arrhythmia Database}

Next we evaluate the proposed CIF algorithm on MIT-BIH arrhythmia database consisting of two-channel ECG records, and hence demonstrate its generalizability. We first optimized the CNN architecture and weights using 5-fold randomized cross validation on training records 200-234 (refer Section \ref{mitbih}) following the steps outlined in Section \ref{sec:ChRes}. The optimized CNN consisted of one filter of length 30 and had no hidden layers, and was used in our CIF heartbeat detector.


\underline{\em Heartbeat detection performance:}
In Table \ref{MITBIHcomp}, we provide comparative performance analysis of the proposed CIF algorithm vis-\`{a}-vis published results. Specifically, we make comparisons in terms of the total numbers of beats, true positives (TP), false positives (FP) and false negatives (FN), as well as sensitivity and PPV values, both gross and average (if available), and the overall score (evaluated using the Challenge formula). We also indicate how many channels each competing algorithm made use of. Notice that the scores, obtained by even the earliest methods, are high. Also, the performance superiority of 2-channel methods over 1-channel methods remains unclear. Those facts possibly indicate that most recorded signals are well behaved. Unsurprisingly, the overall advantage enjoyed by the proposed 2-channel CIF over 1-channel CIF and other competing algorithms was slender, as well. At the same time, compared to 1-channel CIF, the 2-channel CIF reduced FNs from 172 to 72, and increased FPs from 64 to 103, indicating that a recordwise performance comparison could be illuminating. 

\begin{figure*}[ht!]
\centering

\includegraphics[width=18cm]{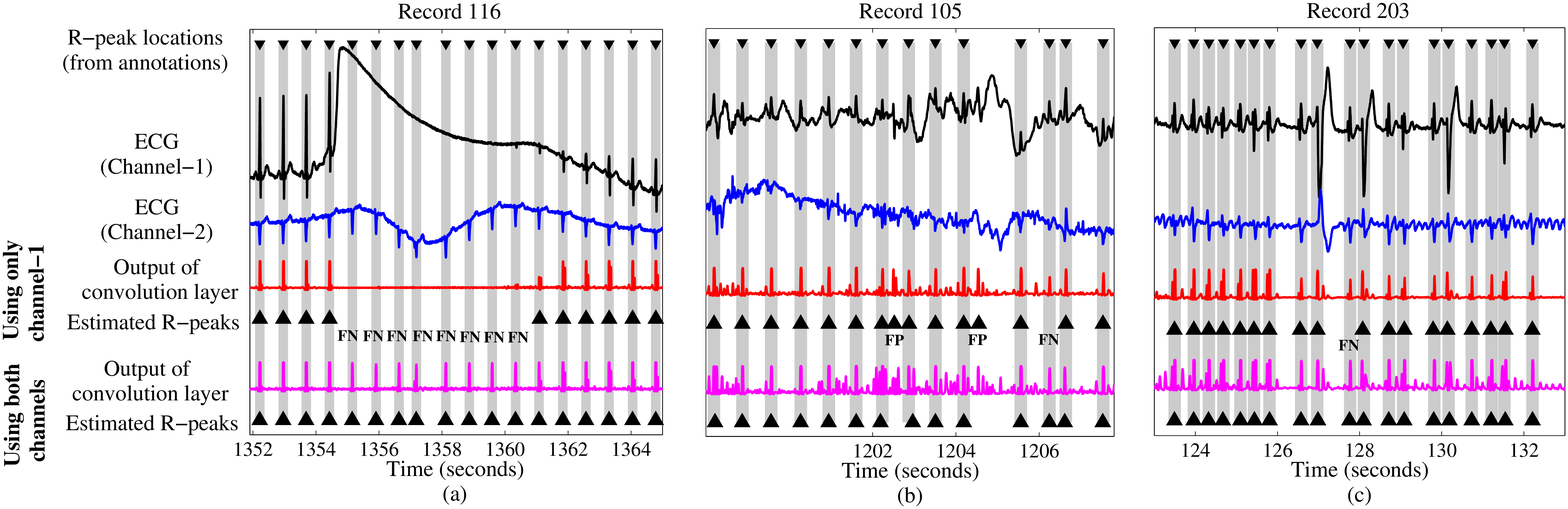}
\includegraphics[width=18cm]{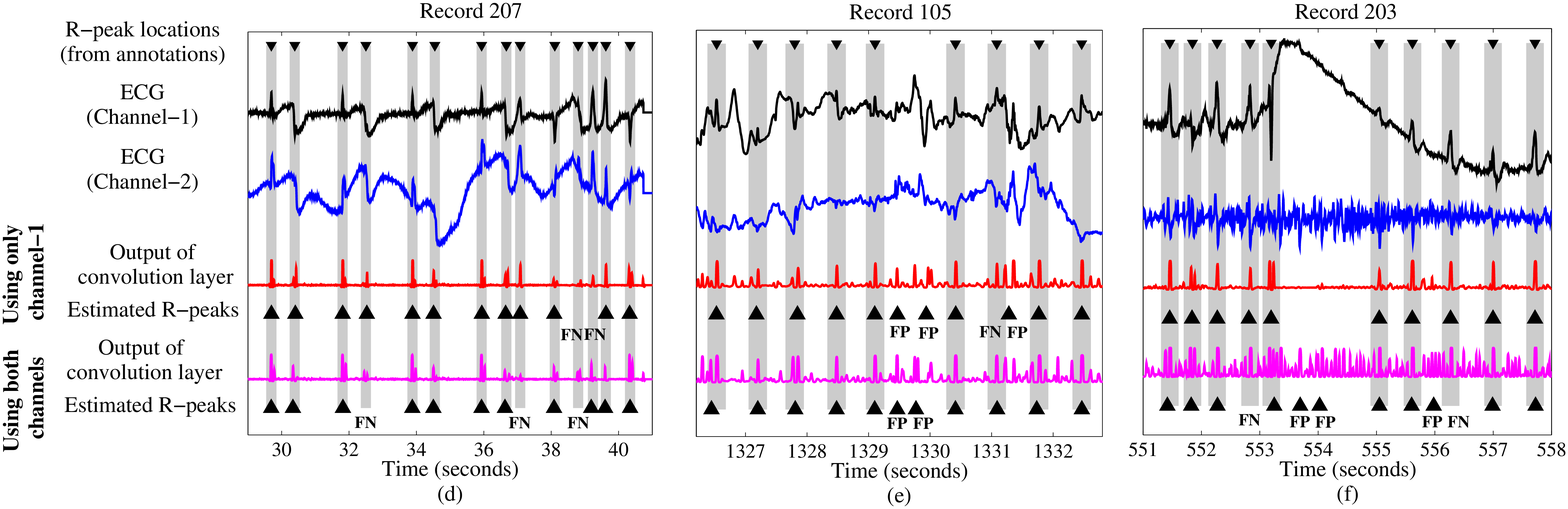}

\caption{Robust heartbeat detection using 2-channel CIF in various cases: (a) missing signal in channel-1, (b) noisy signal, (c) arrhythmic condition; Erroneous detection: (d)-(f) illustrative cases.}
\label{MITBIHdata}
\end{figure*}

Such comparison is presented in Table \ref{FullPerf}, where increase ($\uparrow$), decrease ($\downarrow$) or no change ($-$) in FPs and FNs are indicated. Of the 48 records, 20 see no change at all, and a significant number of those witness minor changes. Let us now inspect the records producing major changes. First, we show in Figures \ref{MITBIHdata}a-c illustrative segments from respective records 116, 105 and 203, where 2-channel CIF corrected errors committed by 1-channel CIF with the help of channel-2. In record 116, FNs incurred using only channel-1 got corrected when channel-2 was also used, because those correspond to episodes of missing signal in channel-1 only. Similar phenomenon also occurs in records 201 and 228 (not shown). In record 105, FPs and FNs incurred by 1-channel CIF got rectified by 2-channel CIF, despite both channels being noisy. In record 203, tagged as `very difficult record, even for humans' \cite{data_desc}, correction of a certain FN is shown. The aforementioned examples demonstrate the robustness of 2-channel CIF in scenarios of varying difficulty levels.  

Despite its general success, our algorithm still fails in certain difficult circumstances. As illustration, we depict in Figures \ref{MITBIHdata}d-f segments from records 207, 105 and 203, where the intended results were not obtained. In record 207, tagged `extremely difficult' \cite{data_desc}, while one FN was corrected, one remained unrectified, and two new ones were introduced, because channel-2 appears to convey confusing information. In record 205, a heartbeat location estimate shifted slightly to correct both an FN and an FP. At the same time, two FPs remained uncorrected due to `high grade noise and artifact' \cite{data_desc}. In the `very difficult' record 203, a shift in location estimate introduced an FP and an FN. Those along side another FN and two FPs resulted because the channel-2 signal, buried in noise, distracted the 2-channel CIF algorithm. In summary, in a variety of circumstances, the proposed 2-channel CIF method tends to correct estimation errors made based on channel-1 using additional information from channel-2, except in certain difficult cases. In the process, while achieving a slightly higher overall score, the 2-channel CIF tended to also bring the four performance indicators, namely, $Se_{avg}$, $PPV_{avg}$, $Se_{gross}$ and $PPV_{gross}$, closer in value to one another, compared to 1-channel CIF.

\section{Conclusion}
\label{conc}

In this paper, we proposed a CNN-based information fusion (CIF) algorithm for heartbeat detection from multiple physiological signals that systematically exploits both temporal and inter-signal dependencies. Specifically, our detector learns a set of linear filters to extract fused features, which are then mapped to estimated heartbeat locations. On the ECG and BP channels of the PhysioNet 2014 Challenge database, our CIF technique achieved a score of 94\%, which is superior to hitherto reported scores. Further, on the MIT-BIH arrhythmia database of 2-channel ECG records, we achieved a score of 99.92\%, which also compares favorably with competing scores. At this point, note that unlike the proposed technique, the existing methods do not generalize to arbitrary databases. For instance, algorithms applied to the Challenge database generally use hand-picked features specific to ECG and BP and do not generalize to 2-channel ECG of the arrhythmia database, and {\em vice versa}. One notable exception to the above is found in an aforementioned algorithm \cite{pangerc2015robust}, which uses hand-picked features but allows upto three channels each of ECG and blood pressure signals, and hence can be applied to the Challenge, MIT/BIH and other databases consisting of ECG and BP records. In contrast, our algorithm is designed to generalize to records consisting of arbitrary physiological signals. The superior performance of our generalizable CIF technique is especially remarkable, because it did not use human knowledge of any signal set, and is based only on features directly learnt from the database at hand. In addition to generalizability and improved accuracy, our method also demonstrated desirable robustness to various clinical anomalies and non-clinical distortions. Accordingly, we envisage CIF-based heartbeat detection being incorporated in medical devices monitoring multiple physiological signals. As a technical improvement, we plan to study the sensitivity-versus-PPV tradeoff by varying the CNN decision threshold (which has been kept constant in this paper), and optimize such threshold subject to operating constraints.

\


%
%
%




\ifCLASSOPTIONcaptionsoff
  \newpage
\fi

\balance



\bibliographystyle{IEEEtran}
\bibliography{Refs_TBME}

\end{document}